\documentclass[12pt, a4paper, parskip, english]{scrartcl}
\usepackage{amssymb}
\usepackage{babel}
\usepackage{lmodern}
\usepackage{amsmath}
\usepackage[T1]{fontenc}    % use 8-bit T1 fonts
\usepackage[%
    backend=biber,
    style=nature,
    url=false,
    doi=false,
    date=year,
    doi=true,
    isbn=false,
]{biblatex}
\usepackage{graphicx}
\usepackage{stackengine}
\usepackage{graphicx}
\graphicspath{ {./figures/} }   % set path for figures
\usepackage{mwe}            % enables placeholder figures
\usepackage{hyperref}       % hyperlinks
\usepackage{url}            % simple URL typesetting
\usepackage{booktabs}       % professional-quality tables
\usepackage{amsfonts}       % blackboard math symbols
\usepackage{microtype}      % microtypography
\usepackage{physics}        % a lot of quality-of-life improvement for math typesetting
\usepackage{csquotes}       % \enquote environment for automatic-language ""
\usepackage[format=plain, font=small, labelfont=bf, labelsep=period]{caption}
\usepackage{subcaption}
\usepackage{cleveref}       % \cref for automatic referencing
\usepackage{xcolor}
\usepackage{authblk}

\usepackage{comment}
\usepackage{hyperref}
\usepackage{cancel}
\usepackage{xr}
\usepackage[colorinlistoftodos]{todonotes}

\title{Entropy-driven decision-making dynamics sheds light on the emergence of the \enquote{paradox of choice}}
\author[a]{Manish Gupta}
\author[b,*]{Arnab Barua}
\author[c,a,*]{Haralampos Hatzikirou}
\affil[a]{Technische Univesit\"at Dresden, Center for Information Services and High Performance Computing, N\"othnitzer Stra{\ss}e 46, P.O. Box: 01062, Dresden, Germany}
\affil[b]{Tata Institute of Fundamental Research, Hyderabad - 500046, India}
\affil[c]{Mathematics Department, Khalifa University, P.O. Box: 127788, Abu Dhabi, UAE}
\affil[*]{Corresponding authors: Arnab Barua, arnabbaruaphysics@gmail.com; Haralampos Hatzikirou, haralampos.hatzikirou@ku.ac.ae}

\begin{document}

\maketitle
Decision making is the cognitive process of selecting a course of action among multiple alternatives. As the decision maker belongs to a complex microenvironment (which contains multiple decision makers), has to make a decision where multiple options are present which often leads to a phenomenon known as the "paradox of choices". The latter refers to the case where too many options can lead to negative outcomes, such as increased uncertainty, decision paralysis, and frustration. Here, we employ an entropy-driven mechanism within a statistical physics framework to explain the premises of the paradox. In turn, we focus on the emergence of a collective "paradox of choice", in the case of interacting decision-making agents, quantified as the decision synchronization time. Our findings reveal a trade-off between synchronization time and the sensing radius, indicating the optimal conditions for information transfer among
group members, which significantly depends  the individual sensitivity parameters. Interestingly, when agents sense their microenvironment in a biased way or their decisions are influenced by their past choices, then the collective "paradox of choice" does not occur. In a nutshell,  our theory offers a  low-dimensional and unified statistical explanation of the "paradox of choice" at the individual and at the collective level.

\vspace{2cm}

\section*{Introduction}

Human decision-making is defined as the result of a thorough evaluation of alternative choices, considering the likelihood and value of the corresponding outcomes \cite{VANDERPLIGT20013309}. Generally,  decision-making  in organisms ranges from individual choices influenced by past experiences, perception, and knowledge, to collective decisions manifested in collective dynamics \cite{MacLean2012}. This spans from the microscopic movements of cells \cite{Mehes2014} to macroscopic behaviors observed in flocks of birds and schools of fish \cite{deutsch_collective_2012}. Similarly, human make individual decisions which by virtue of interaction lead to emergent societal changes \cite{Nava-Sedeno2020}. In these processes, individual decision-makers integrate the information produced by nearby peers, through sensing and interactions \cite{VICSEK201271}, demonstrating how individual actions can aggregate to produce complex collective dynamics. 

The presence of extensive information in the local environment introduces a paradoxical situation, famously known as the \textit{\enquote{paradox of choice}}. Coined by psychologist Barry Schwartz, this paradox encompasses decision paralysis, dissatisfaction with choices made, and an overall decrease in the well-being of decision-makers \cite{Schwartz2007}. A critical question arises: how much information is \textit{too much}? As environments have finite sizes, determining the optimal amount of information necessary for effective global decision-making at the group level becomes crucial. To quantify this situtation, British and American psychologists William Edmund Hick and Ray Hyman experimented on this and found that there exists a crucial relationship between the average reaction time of each decision-maker and the number of choices\cite{Hicks1952, Ray1953}. This law is known as \textit{Hick's Law} and states that the average reaction time is a logarithmic function of a number of options. the decision-making process becomes more complex and time-consuming, which can lead to frustration, being overwhelmed with information, and decreased satisfaction at the personal level. To understand the \enquote{paradox of choice}, we employ the \textit{LEUP} theory or Least Environmental Uncertainty Principle \cite{Hatzikirou+2018} , which posits that individual agents or organisms strive to reduce their microenvironmental entropy over time as they assimilate knowledge from their surroundings. This principle has been applied to various biological systems such as cell migration \cite{Barua1}, Epithelial-Mesenchymal-Transition \cite{Barua2}, and cell differentiation \cite{e23070867}.

In this paper, we first define an internal variable reflecting individual choices, driven by the entropic gradient within the microenvironment, and hypothesize that discontent arises as the variance/uncertainty of this variable increases over time. Using mean-field calculations, we derive a formula quantifying the variance rate relative to the number of choices, revealing an optimal number of choices that minimizes this rate. We then analyze decision-making dynamics using a Metropolis-Hastings Monte Carlo method within an individual-based model, identifying a trade-off between convergence time and sensing radius that optimizes information transfer among group members based on individual sensitivity parameters. Furthermore, we explore weighted decision-making influenced by the Von-Mises distribution, showing that this approach can accelerate group decisions depending on weight variance, and investigate trend-setting behaviors leading to micro-level community formation. Comparing our findings with the \textit{Vicsek Model}\cite{Vicsek1995}, we show that low weight variance yields similar outcomes, while high variance offers insights into the paradox of choice.

\section*{Mathematical modeling of interacting  decision makers}

%\textcolor{blue}{\subsection*{}}
Moving and interacting agents are modeled by a two-dimensional self-propelled particle model (SPP). In this model, N $\in \mathbb{N}$ cells move on a two-dimensional space. The n-th agent is characterized by its position, $\mathbf{r}_n \in \mathbb{R}^2$, speed, $v_n \in [0, \infty) \subset \mathbb{R} $, and an orientation $\theta_n \in [0, 2\pi) \subset \mathbb{R} $.  Here for simplicity, we assume that all the agents are moving with the same constant speed with time. 
The change in orientation of each agent results from the interaction potential $U(\mathbf{r}_m,\theta_m)$, which depends on the positions and orientation of other agents within the interaction radius $R\in\mathbb{R}_+$. This interaction potential dictating the dynamics of orientation is specified using the LEUP theory that states an individual updates its decision in such as way that it minimizes the entropy of its micro-environment, i.e.~$S(\Theta|\theta_n)$ where $\Theta=\{\theta_m:||\mathbf{r}_m-\mathbf{r}_n||\leq R\}$ the collection of microenvironmental orientations.
Assuming this the rate of change in the orientation can be expressed as depicted in Eq.(\ref{langevin}). The microenvironmental entropy $S(\Theta|\theta_n)$ term acts as interaction potential. Please note that we calculate microenvironmental entropy numerically using the histogram to obtain the probability distribution of the orientations of neighbourhood decision makers.  The responsiveness of the agent to follow the potential gradients is regulated by the parameter $\beta \in \mathbb{R}$, called angular \textit{sensitivity}. This parameter is associated with the sensed \textit{information processing} since it leads to decisions that either decrease ($\beta>0$) or increase ($\beta<0$) microenvironmental entropy.
Additionally, orientation fluctuations occur due to stochastic noise terms $\xi_{n}(t)$ where  $ t \in \mathbb{R}_{\+}$ denotes time. The noise $\xi_{n}(t)$ will be assumed to follow a normal distribution, which has the statistical properties $\langle\xi_{n}\left(t\right)\rangle=0$ and $\langle\xi_{n}\left(t_{1}\right)\xi_{m}\left(t_{2}\right)\rangle=2D\delta\left(t_{1} - t_{2}\right)\delta_{nm}$, where $t_{1}$ and $t_{2}$ are two-time points.
\begin{equation}
  \frac{d\theta_n}{dt}=-\beta \frac{\partial S(\Theta|\theta_n)}{\partial \theta_{n}} + \xi_n(t)
  \label{langevin}
\end{equation}
Further, we assume that the decision-making process is at a steady state. Accordingly, the steady-state probability distribution of the orientation of n-th agent is given as,
\begin{equation}
  P(\theta_n)\propto e^{-\beta S(\Theta|\theta_n)}
\end{equation}
The decision-making step for an agent is initialized by randomly choosing an orientation from its micro-environment and proposing this to be its new orientation state. The \textit{Metropolis-Hastings} algorithm will decide whether to accept or reject this proposed state. If the choice results in a decrease of the microenvironmental entropy $\Delta S <0$, the proposed state is accepted. If the choice results in an increase of the entropy $\Delta S>0$ then the proposed state is accepted with probability $e^{-\beta \Delta S}$, where $\Delta S$ is the change of the entropy before and after the update. These steps are repeated and the system is allowed to reach steady state, after which the properties of the system are measured. Now, Let us define the normalized complex orientation of the n-th cell, $z_n \in C $ \text{as}  $z_n = e^{i\theta_{n}}$, where $i$ is the imaginary unit. The k-th moment of the orienation over an area A is given by $\langle z_k \rangle_{A} = \frac{1}{N_{A}}\Sigma_{m\in A} z^{k_{m}} $, where the sum is over all cells in area $A$, and $N_{A}$ is the total number of cells in $A$. The polar order parameter in the area $A$ is given by
\begin{equation}
  {S_{A}}^{1} = \mid\langle z\rangle_{A}\mid  
\end{equation}
Based on the polar order parameter, we define the group level decision making as the collective orientation. Please note that, we define individual agent's decision as the orientation $\theta_{n}$. So in the macrocopic level when the order parameter reaches a steady state we define that a group level decision has been made. When system agents make a flock then the polar or global order parameter converges to the value $1$. We characterise this value as the maximally learned state as they are ordered and the for the whole system microenvironmental entropy reaches close to $0$ or the information gain is maximum. On the other hand, for the random or disorder macroscopic state  the ${S_{A}}^{1}=0$, where the for the  system microenvironmental entropy increases to a particular maximal value or the information gain is minimum. The decision making time at the group level is defined by the synchronization time $(\tau)$ at which the polar order parameter reaches $0.9$ value.  In the next section, we have analyze our model based on the variations two parameters i.e,~(\textbf{i}) polar order parameter and (\textbf{ii}) the synchronization time. Later, we include a finite memory (i.e., an array of orientations stored of memory capacity $m$ depicted by last $m$ time steps) for each decision maker to provide further realism in the decision making process. Finally, we compare our model results to the standard Vicsek model in the zero noise case where the decision on angles of the $n$-th decision maker at the next time step $t+\Delta t$ is given by the average angular decision in the microenvironment (i.e., $\langle\Theta(t)_r\rangle$) in the previous time $t$ step \cite{vicsek1995novel}.
\begin{equation}
    \theta_{n}(t+\Delta t)= \langle\Theta(t)_r\rangle,
\end{equation}
where $\langle\Theta(t)_r\rangle=\tan^{-1} (\langle \sin(\Theta)\rangle/\langle \cos(\Theta)\rangle$ and $\Delta t$ the corresponding time step.

\section*{The  \enquote{paradox of choice} as a minimum in the decision uncertainty rate for individual decision-makers }
 % \section*{Mean field result of our model}
In this section, we attempt to understand why we can observe the \enquote{paradox of choice} in individual decision-making using LEUP dynamics. Here, we model decision-maker's satisfaction with  the uncertainty rate of the decision (internal) variable when the number of choices is fixed.

Let us consider a microenvironment consisting of total $N$ decision-makers, focusing on the $n$-th decision-maker at the center. At time $t$, this decision-maker samples all microenvironmental decisions in a Gaussian manner, where the mean value of the collected decisions is defined as $\mu_{n}$ and the sample variance is denoted as $\sigma_{n}^{2}$. For simplicity we define the angular decisions lies on the interval $(0,2\pi)$. To evaluate the average decision-making at the group level, we adopt a mean-field approach, utilizing the Central Limit Theorem (CLT) to obtain the scaling of the quantities of interest with $N$. Assuming the Gaussian form of the microenvironmental entropy $S(\Theta|\theta_n)=\frac{1}{2}\ln{(2\pi e \sigma_n^2)}$ and using Eq.(\ref{langevin}), we can express the rate of change of decisions as follows:
\begin{equation}\label{eq:LEUP}
    \frac{d\theta_n}{dt}=-\frac{\beta}{\sigma_{n}^{2}}\left({\frac{\theta_n}{N} -\frac{\sum_{i\neq n}^{N} \theta_i}{N^2}}\right) +\xi_n(t)=-\beta f+\xi_n(t)
\end{equation}  
where $\xi_n(t)$ is a noise following Normal distribution $\mathcal{N}(0,D)$ and the force term $f$ reads 
\begin{equation}\label{eq:force}
    f=\frac{1}{\sigma_{n}^{2}}\left({\frac{\theta_n}{N} -\frac{\sum_{i\neq n}^{N} \theta_i}{N^2}}\right) 
\end{equation}
Now, as we already consider that the decision making occurs inside a local environment close to the steady state, one can use a mean-field  approach to illustrate the temporal evolution of variance for each average individual. Using the previous dynamical equation of decisions, one can write the dynamics of variance \cite{Oku2018}: 
\begin{equation}\label{variance}
\frac{d\Sigma}{dt}=D-2\beta\Sigma\frac{\partial f}{\partial \theta_n}=D-2\beta\Sigma R 
\end{equation}
The macroscopic variance of the random variable has been defined by $\Sigma=\frac{N}{N+1}\sigma_n^2$ (for details see SI) and the variance of the Gaussian noise is define by $D$ and the $R$ drift term reads as
\begin{equation}
      \frac{df}{d\theta_n}=\frac{1}{\sigma^{2}_{n} N}-\frac{\theta_n}{N (\sigma_{n}^2)^2}\frac{d\sigma^{2}_{n}}{d\theta_n}+\frac{\mu_n}{N (\sigma_{n}^2)^2}\frac{d\sigma^{2}_{n}}{d\theta_n}-\frac{\theta_n}{(\sigma_{n}^2)^2N^2}\frac{d\sigma^{2}_{n}}{d\theta_n}
\end{equation}
Using the above expression of $\frac{df}{d\theta_n}$, one can find the rate of the change of variance over time $t$ with respect to number of the agents in the environment $N$ and sensitivity $\beta$ as (please see SI for details)
\begin{equation}
\frac{d\Sigma}{dt}=-2\beta R(\Sigma,\mu,N) \Sigma
\end{equation}
\begin{equation}
\frac{d\Sigma}{dt}=-\frac{2\beta}{(1+ N)}+\frac{4\beta}{(1+N)^2}+
    \frac{8\beta}{N(1+N)^2}=-\frac{2\beta\left(N^2-N-4\right)}{N(1+N)^2}
\end{equation}

For negative $\beta$, the variance will always increase over time as the corresponding rate is positive, converging to zero. However, for $\beta>0$, the variance rate's dependence on $N$ is non-monotonic, exhibiting a minimum. This suggests that the decision-maker experiences increased satisfaction with a small increase in the number of options $N$, but beyond a certain point, the \enquote{paradox of choice} emerges.  According to our theory, a low $N$ results in poor sampling of the microenvironment, leading to unsatisfactory decision-making. While increasing $N$ initially improves decision-making to an optimal value, surpassing this point causes the variance to scale increasingly with a larger number of options. In the following, we will explore the emergence of this paradox during collective migration.

%Hence we expect the synchronization time will also have the non monotonicity with N. As we see in the result of our numerical simulation that synchronization time is non monotonically dependent on interaction radius (R). If we consider the homogenous distribution of particles then N $\propto$ $R^2$.

\begin{figure}[h!]
    \centering
    \includegraphics[width=0.7\linewidth]{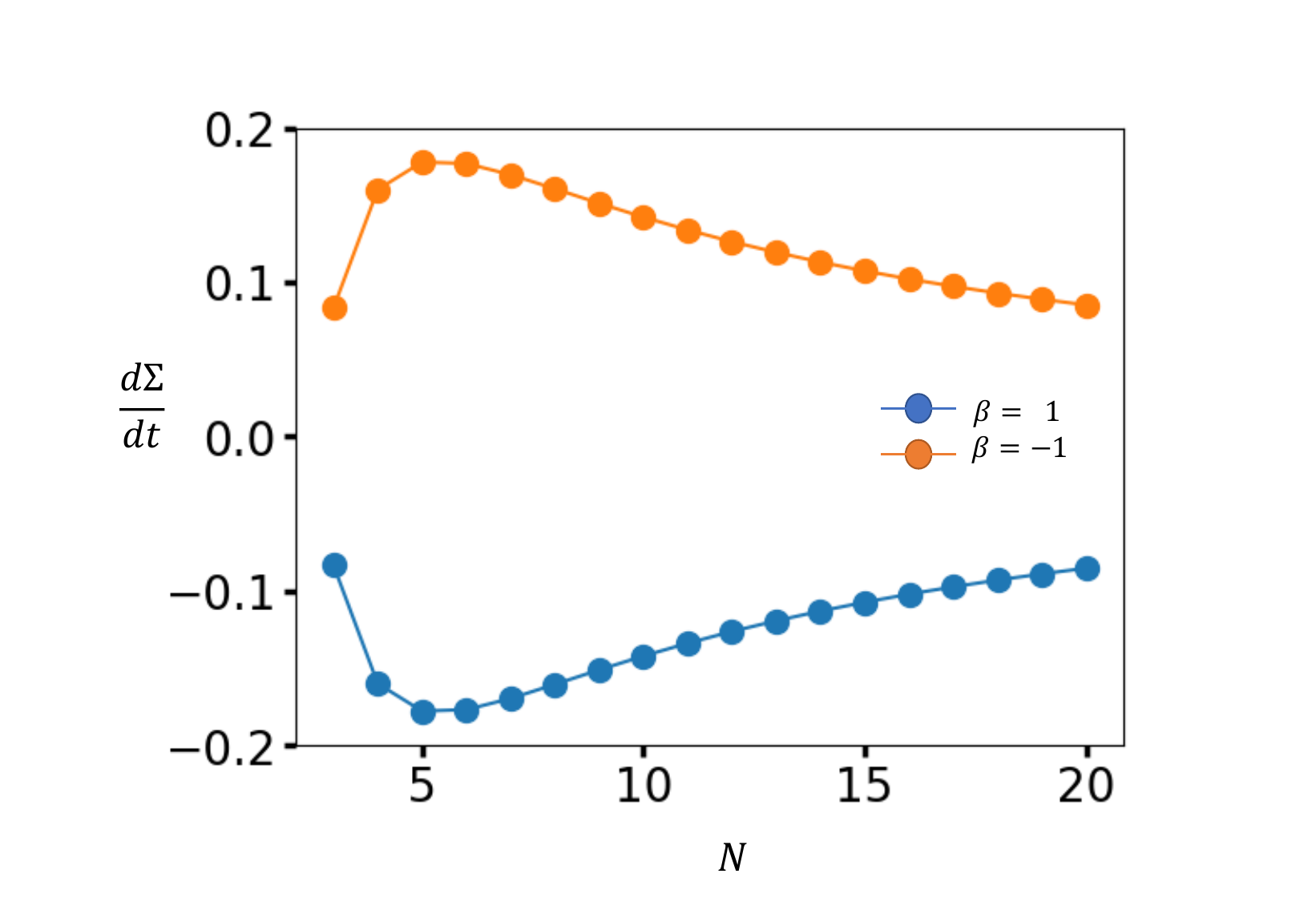}
    \caption{Mean field result for the rate of change in the variance of decision x in the system vs number of agents (N) in the microenvironment. }
    \label{density}
\end{figure}

Another manifestation of the \enquote{paradox of choice} could be measured via the convergence time to equilibrium $\tau$, just like in Hick's law. In particular, we can calculate that the individual state dynamics follows the eq.~(\ref{eq:LEUP}). The factor next to the drift term denotes the relaxation rate to equilibrium, implying that $$\tau\propto \frac{\sigma_n^2}{\beta}.$$ This suggests that the decision uncertainty increases the relaxation time of the state variable towards its equilibrium value. The latter observable will be used to assess the emergence of the collective \enquote{paradox of choice}. 

\section*{The emergence of the collective \enquote{paradox of choice}}

We postulate that the \enquote{paradox of choice} is an environment-driven phenomenon, where individuals make decisions in response to their surroundings. In this section, we aim to understand how a system of individual decision-makers can arrive at faster group-level decisions despite the presence of this paradox. In particular, we will focus on the synchronization time where in the mean-field should coincide with the mean orientation relaxation time. It is expected a minimum synchronization time for an increasing sensing radius, and fixed sensitivity parameter, will be interpreted as a collective manifestation of the paradox.

\subsection*{Interplay between interaction radius and the sensitivity in the dynamics of polar order}

Sensing the microenvironment is the first step in decision-making. In a multiagent microenvironments, agents sense and interact with other agent states within an interaction radius $R$. A large interaction radius will provide the agent with a better picture of the microenvironment, since the number of agents $N\propto R^d$ where $d$ is the system dimension. At the same time sampling noise scales with number of agents as $\sim\sqrt{N}$. This raises the question: does it exist an optimal sensing radius? Once an agent has information about the microenvironment, the next step is to process that information to make a decision, i.e., altering its internal state. Here, we assume that the underlying principle for decision-making is the Least Environmental Uncertainty Principle (LEUP). As discussed in the methods section, the sensitivity parameter \(\beta\) associated with this decision-making principle controls the response of a decision-maker and the type of information processing. For instance, if the sensitivity is \(\beta = 0\), then decisions are rendered randomly. However, higher \(\beta\) indicates increased responsiveness of the agent to the environment. Additionally, one can think of the sensitivity parameter \(\beta\) as a coupling parameter that can help the total system reach a synchronized or desynchronized state \cite{strogatz}. 

\begin{figure}[ht!]
    \centering
    \includegraphics[width=0.8\linewidth]{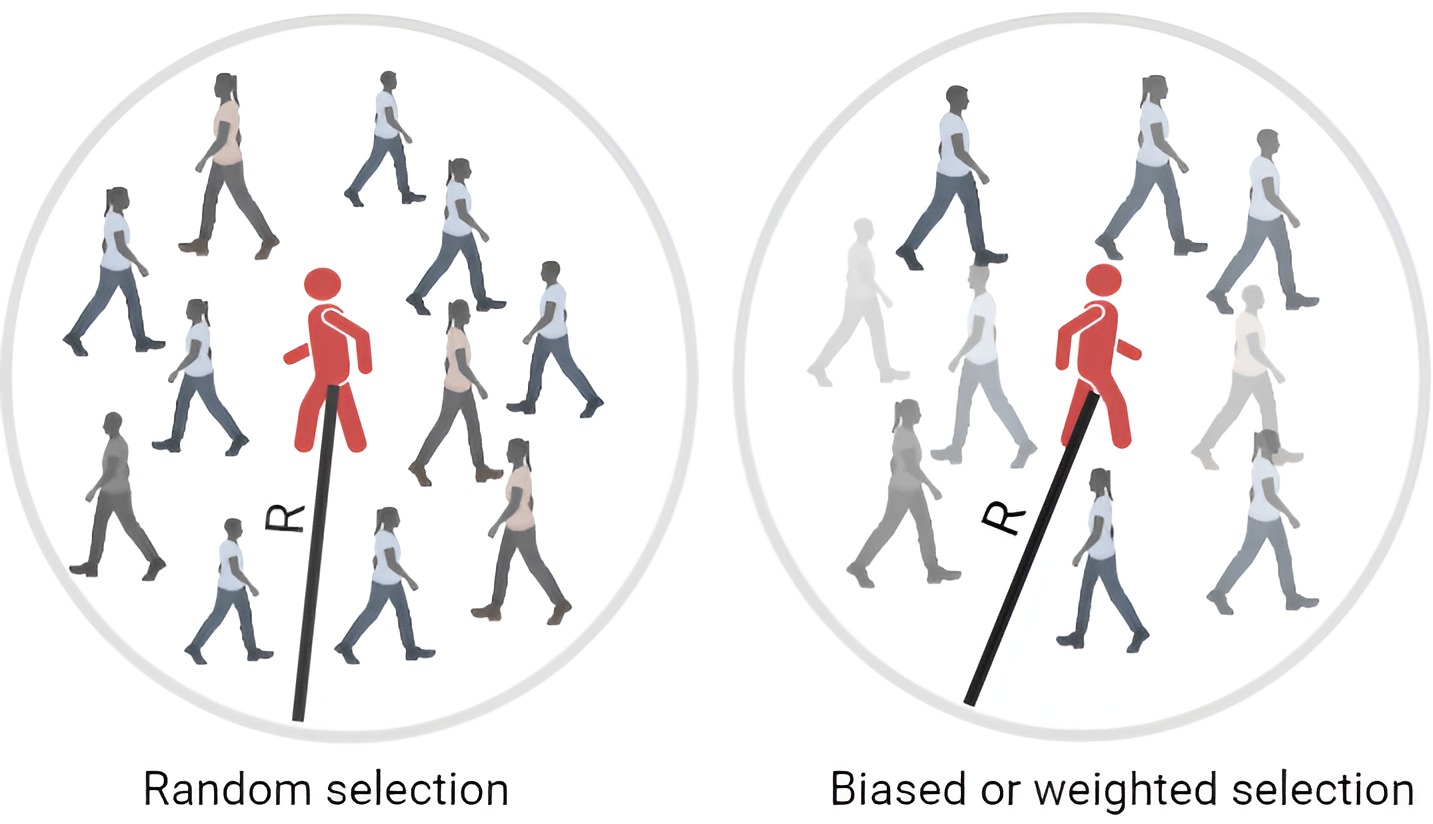}
    \caption{Figuring out decisions based on random selection and biased or weighted selection through a schematic image. In the weighted selection case, the sampling of neighboring agents is biased by the mean orientation of the interaction neighborhood. Here $R$ is the sensing radius.}
    \label{sch1}
\end{figure}

To this end, we investigated the interplay between interaction radius (\(R\)) and sensitivity (\(\beta\)) on the synchronization of the system for three different cases. The radius $R$ controls the capacity of the options (agents) number. These cases differ in the level of selection of one of the states from the microenvironment. In the  case of \textit{random sampling}, an agent will randomly sample a state from the microenvironment and then, according to the entropy based Monte Carlo rule, the orientation state will be accepted or rejected. However, in realistic scenarios, agents do not randomly sample their microenvironment; instead, they sense it in a weighted manner. Here agents use a \textit{biased sampling}, where an agent samples the interaction neighborhood using a Von Mises distribution mean the corresponding to the average orientation in the microenvironment (for an illustration of the sampling methods see Fig.~(\ref{sch1}). We study both cases: (i) random and (ii) biased/weighted sampling. Later, we investigate the role of trends in group-level decision-making in the (iii) effect of decision memory on the "paradox of choice".

\subsubsection*{Random sampling}

As mentioned above,  in our work, agent decision-making  is governed by the LEUP. According to Eq. (1), setting \(\beta < 0\) results in decisions that maximize the microenvironmental entropy. As expected, maximizing entropy at the mesoscopic scale fails in leading to global synchronization. Conversely, minimizing entropy at the mesoscopic scale (i.e., \(\beta > 0\)) leads to synchronization beyond a threshold interaction radius. The existence of this threshold radius, beyond which synchronization occurs, is consistent with the previously studied multi-decision-maker model \cite{Barua1}. Agents need a minimum level of interaction to synchronize their decisions. For a radius below this threshold, the number of particles in the neighborhood is sparse, causing weak interaction and, hence, failure to synchronize. So far, we have shown the conditions necessary for the system's synchronization. Now, we will investigate the impact of \(\beta\) and \(R\) on the synchronization time \(\tau\). Synchronization time is defined as the time taken by the decision-makers in the system to synchronize their orientation i.e.,~the global polar order reaches 0.9.

\begin{figure}[ht!]
    \centering
    \includegraphics[width=1\linewidth]{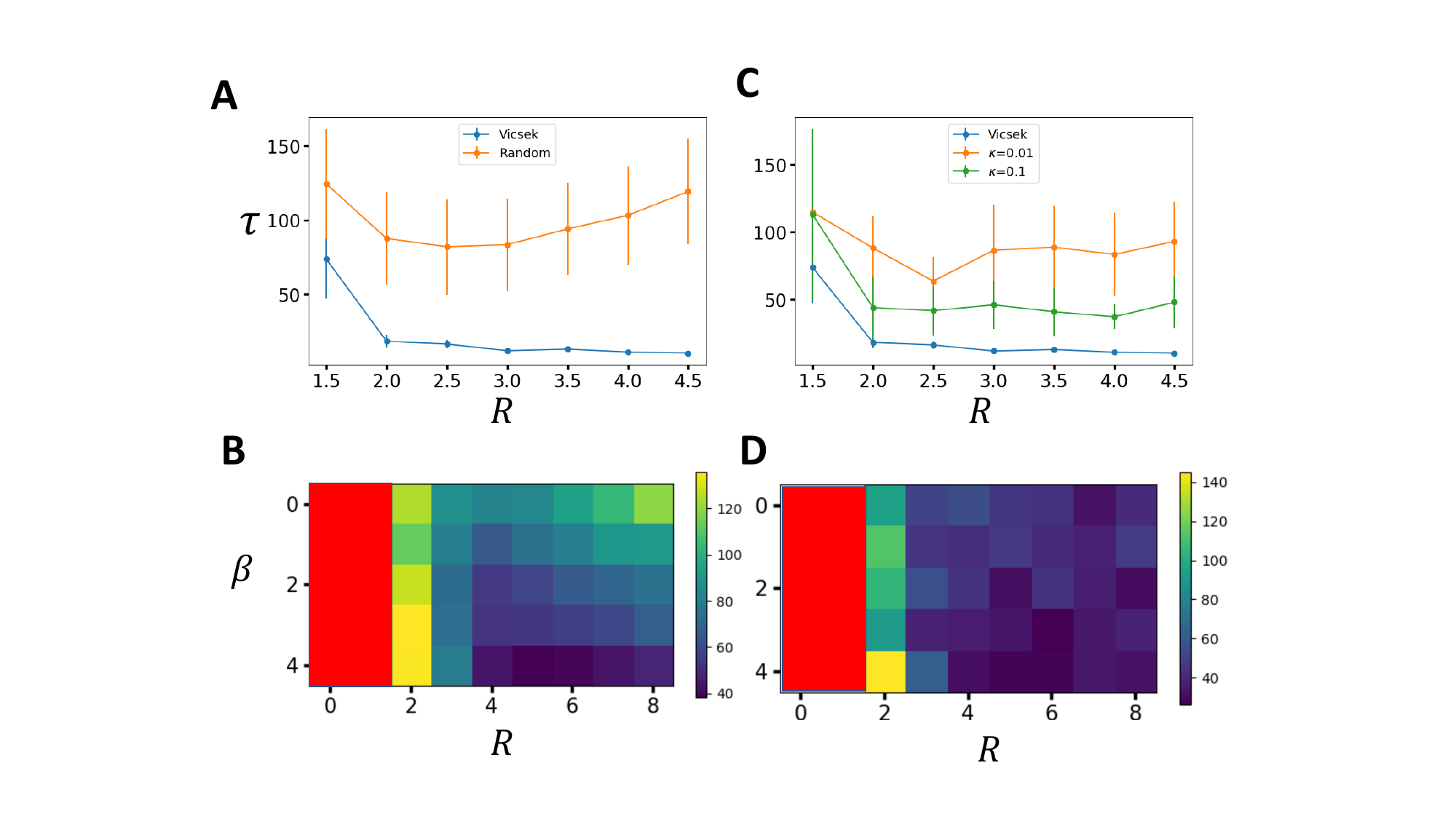}
    \caption{\textbf{Synchronization time vs interaction radius} (A) and (C) represent the mean synchronization time of the system for 20 simulations on varying sensing radius for random ($\beta = 2$) and Von Mises distribution respectively. The comparison of synchronization time with the \textit{Vicsek Model} is also shown for each figure. (B) and (D) shows the phase diagram with synchronization time averaged over 20 simulations, plotted against $\beta$ and R for random and Von Mises distributions ($\kappa =0.1$) respectively. Synchronization is defined by achieving a global polar order greater than 0.9. The red region in the phase diagram corresponds to the regime where the system fails to synchronize. All the simulations were done for the speed $0.1 \,unit/sec$}
    \label{heatmap}
\end{figure}

In the Vicsek model, an increase in interaction radius decreases the synchronization time, as shown in Fig. (\ref{heatmap} A). Interestingly, in contrast to the Vicsek model, our model shows that there is an optimal interaction radius for a given \(\beta\), at which \(\tau\) is minimized, as shown in Fig. (\ref{heatmap} A). This phenomenon is counter-intuitive, as we would expect that increasing the radius enhances interactions, thereby decreasing synchronization time. This behavior is consistent with the "paradox of choice", where an agent exposed to too much information becomes paralyzed, leading to deterioration in decision-making.

Furthermore, we investigate how this decision paradox is affected by sensitivity \(\beta\). To understand this, we plotted a phase diagram, shown in Fig. (\ref{heatmap}B), which represents the relationship of the synchronization time with respect to \(\beta\) and \(R\). Along the \(\beta\) axis, the synchronization time always decreases, implying that increasing sensitivity for a fixed \(R\) enhances the system's decision rate. As shown above, for a fixed \(\beta\), the synchronization time first decreases and then increases, indicating that increasing the amount of information does not necessarily improve decision making. Most importantly, the synchronization time always decreases along the left-to-right diagonal. Thus, if an agent acquires more information, it needs to increase its sensitivity as well to maintain the rate of synchronization.

This finding reveals a trade-off in collective decision-making that strictly depends on sensitivity and interaction radius. We show that decision making is indeed a complex process that relies not only on the volume of information available but also on the individual information processing as quantified by the sensitivity parameter $\beta$. When a certain level of sensitivity is present, an overload of information can impair decision-making, a phenomenon often referred to as the paradox of choice. This highlights the importance of balancing information intake and processing to make effective decisions, as shown in Fig.(\ref{sch1}).

\subsubsection*{Biased  sampling}
Now we employ the biased or weighted sampling scenario for agent decision-making. We calculated the weight for each possible decision using the \textit{Von-Mises} distribution, which is equivalent to the normal distribution for any wrapped random variables. The mean of the distribution is set as the average orientation of the agents in the neighborhood, representing the maximum likelihood estimation of the \textit{Von-Mises} distribution \cite{Abeyasekera1982}. The distribution parameter \(\kappa = \frac{1}{\sigma^2}\) where $\sigma$ is the variance of the sampled neighborhood.

One might wonder how weighted decision-making can impact synchronization. Our simulation shows that for lower value of $\kappa$ implying weak bias, the dependence of synchronization time on interaction radius exhibits similar behavior to the random selection case. As shown in Fig.~(\ref{heatmap} C), the synchronization time decreases to \(R=2.5\), after which it starts to increase. Therefore, the \enquote{paradox of choice} emerges in this case.

For higher $\kappa$, corresponding to strong bias, we observe a markedly different behavior. As shown in Fig.(\ref{heatmap} C), biased agents do not suffer from information overload due to an increased interaction radius. The excess information does not seem to significantly change synchronization time. This observation is qualitatively similar to what is seen in the \textit{Vicsek Model}. This indicates that entropy-driven decision-making, when combined with some additional knowledge, reproduces the results of the \textit{Vicsek Model}. Furthermore, we demonstrate that the paradox of choice, when presented with numerous options, can be mitigated by educating decision-makers with prior knowledge.

\subsubsection*{Memory effect}
So far, we have assumed that agents do not remember any of their past decisions and hence are memoryless. However, in reality, a decision maker's choice is influenced by previous decisions. It is intriguing to study how the implementation of memory alters synchronization in the system. Does the introduction of memory change the nature of \enquote{paradox of choice}?

\begin{figure}[ht!]
    \centering
    \includegraphics[width=0.7\linewidth]{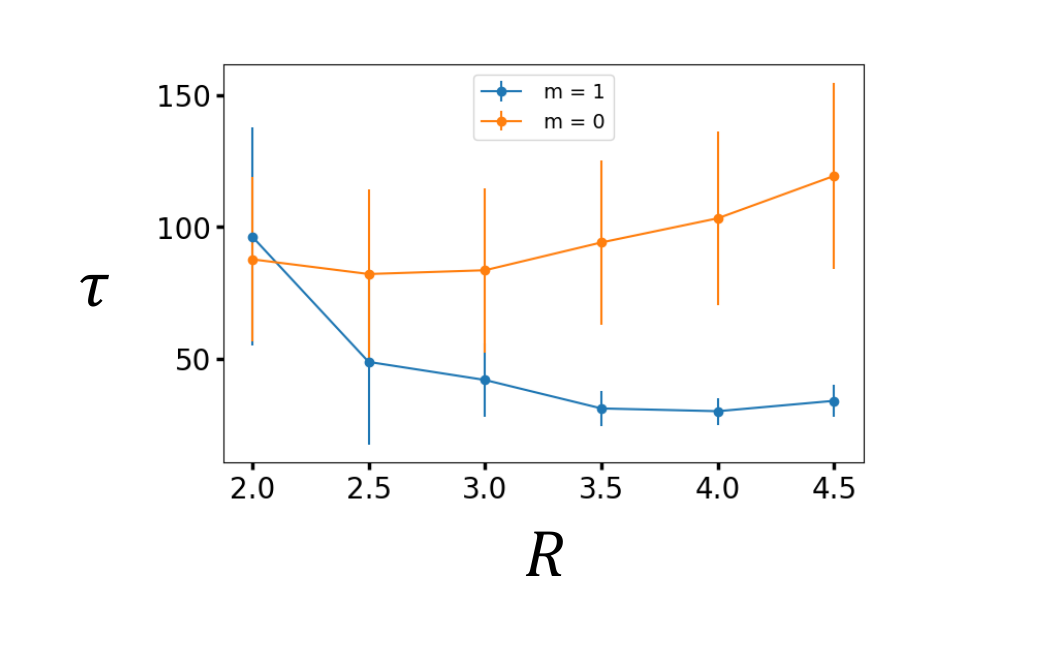}
    \caption{\textbf{Memory effect on synchronization} for agents with memory capacity m =1 and without memory. The simulation is done for $\beta =2$ and 250 agents moving at the speed 0.1 unit/sec.}
    \label{memory}
\end{figure}

We incorporate the function of memory by adjusting decisions based on the surrounding environment, allowing memory to motivate agents to choose decisions that have been previously selected. Each agent possesses restricted memory, meaning that they can only recall their last \(m\) decisions with equal intensity. Interestingly, we discovered that an excess of information does not hinder the agent decision-making process. In fact, agents become more proficient when they possess memory compared to when they are devoid of memory(\ref{memory}). This implies that the paradox of choice is hindered in the case of decision-makers with memory.

%\newpage

\section*{Discussion}

In this article, we attempted to provide an alternative explanation of the \enquote{paradox of choice} based on statistical mechanics arguments. In particular, we have assumed that agents are employing the LEUP as a mechanism of decision-making. According to LEUP, we have defined an internal variable that reflects the choice/state of the individual. The corresponding dynamics is driven by an entropic gradient of choices found in the respective microenvironment. In order to explain the \enquote{paradox of choice}, we made the assumption that individual feel discontent when the uncertainty of this internal variable increases in time.  Using a mean-field calculation, we provide a closed formula that quantifies the rate of variance against the number of choices. Even in this simplifying setting, we find an optimum for a number number of choices that minimizes the variance rate of the intrinsic variable. Above the optimal value of $N$, the decision-maker experiences the frustration due to the overwhelming amount of choices. On the other hand, when number of  choices are too few, then decisions are almost random since the deterministic entropic drift is very weak.
\begin{figure}[ht!]
    \centering
    \includegraphics[width=0.8\linewidth]{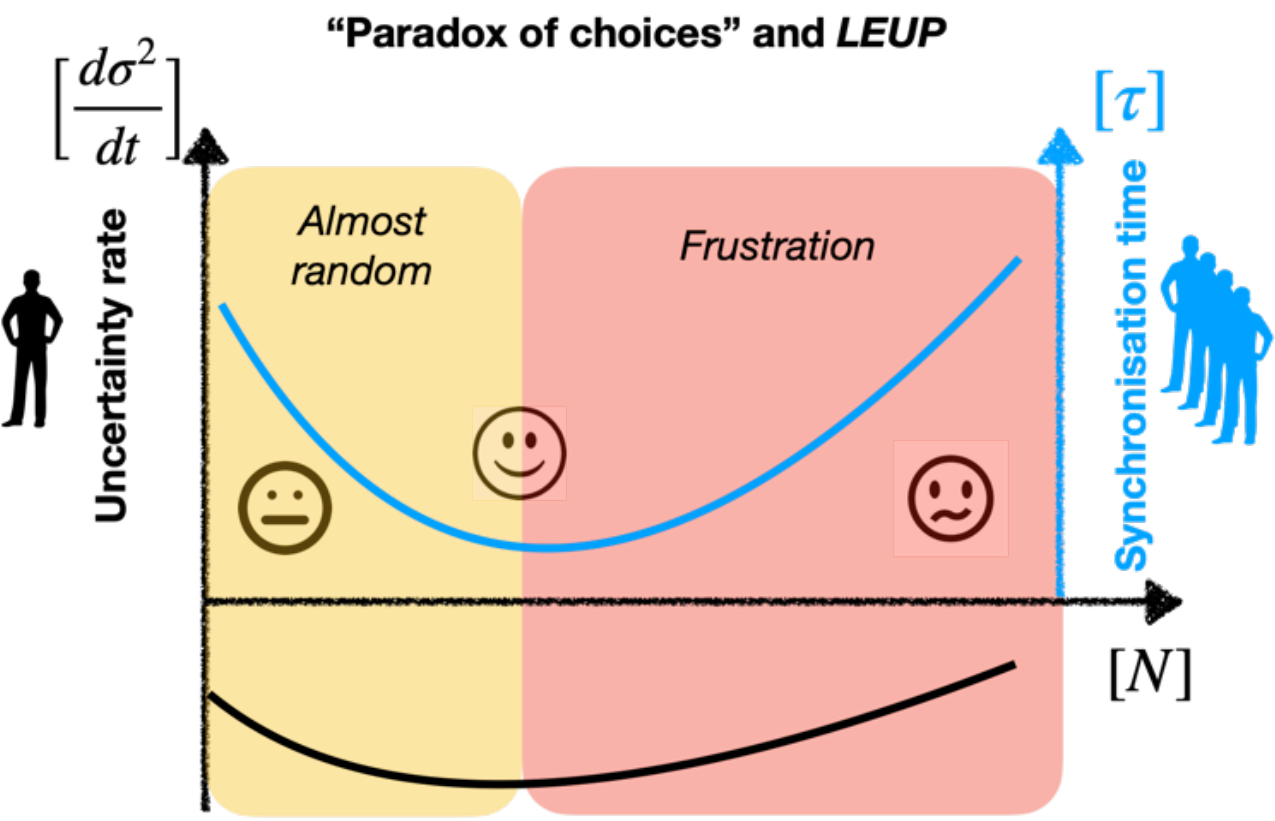}
    \caption{A schematic way understand how our theory contributes to the understanding of the \enquote{paradox of choice}. On the left vertical axis, we indicate the relevant observable regarding individual decision making being the decision uncertainty rate $\frac{d\sigma^2}{dt}$. On the right, with blue, we depict the collective variable being the synchronization time $\tau$. The x-axis represents the number of choices $N$. More information in the text.}
    \label{fig:explaination}
\end{figure}
In turn, we analyzed the microenvironmental entropy of orientation  in decision-makers, using  an individualized-based model to study group-level decision-making dynamics. Our findings reveal a trade-off between synchronization time and the sensing radius, indicating the optimal conditions for information transfer among group members, which significantly depends on the individual information processing (sensitivity parameter). Fig.~(\ref{fig:explaination}) illustrates the main conclusions of our study.

In realistic scenarios, decision-makers observe and respond to their local environments not randomly but in a weighted manner, potentially reinterpreting the \enquote{paradox of choice}. By assigning weightages visualized from the \textit{Von-Mises} distribution, decision-makers realistically adapt to the orientations of local agents, leading to shifts in critical timeframes for decision-making. These shifts suggest that weighted decision-making processes can accelerate group-level decisions depending on the variance of the assigned weights. Moreover, we explore the phenomenon of trend-setting among decision-makers, akin to the \textit{Hipsters effect}, where anti-conformists seeking uniqueness paradoxically end up appearing similar \cite{Touboul2019}. This setup led us to discover the emergence of micro-level communities based on orientations and delays in group decision-making as memory sizes increased.

This interpretation of \enquote{paradox of choice} might be relevant for Cell Biology. LEUP theory has been developed in the context of cell decision-making in multicellular systems \cite{Hatzikirou+2018,Barua1,Barua2,Barua2023}. Cell fate determination is a cell decision-making mechanism related to cell differentiation and development. Stem cells differentiated to more specified phenotypes when they find themselves in well-defined niches \cite{Hicks2023}. Stem cell niches are composed by  a specific variety of relevant determinants, such as nutrients, chemical cues, ECM molecules, biophysical conditions and other cell types, that drive stem cells to acquire a specific phenotype. Each tissue has a uniquely defined niche. Having to few determinants stem cells are not able to initiate proliferation and/or differentiation programs. On the other hand, having more components has lead to impaired differentiation e.g.~by excessive signaling and oversaturation of receptors,  overcrowding of niche cells leading to low proliferative potential or excess ECM components putting physical barriers to differentiation processes. Regarding cell fate decisions, we postulate that there is a trade-off between the information processing capacities of the stem cell and the available microenvironmental information \cite{SU2022408}.

In our model, we assumed that each individual decision maker seeks to minimize their microenvironment entropy, treating it as a cost function that we computed numerically. It will be interesting to explore how our model can be applied to various systems as a utility or cost function with different constraint setups, which can be validated using experimental evidence. In addition, we can observe the behavior of decision makers under these conditions. This framework is crucial not only for understanding paradoxes, but also for comprehending the macroscopic phenomena of group-level decision-making based on local influence. Upon further consideration, we can also evnision the application of this framework within the context of opinion dynamics or voting dynamics, where individual agents share their opinions in a closed community, potentially leading to normative social influence driven by conformity \cite{RevModPhys.81.591}. When the order parameter is close to 1, it can indicate a unique group-level decision, reflecting a democratic consensus.

In future work, this framework can serve as a benchmark or training model (similar to the bootstrap method), where individual agents train their sensing radius and sensitivity based on the local density of microenvironmental agents and neighborhood dynamics. This trained model can then be applied to new environmental setups, allowing individual agents, as a group, to enhance decision-making time at the group level based on different utilities or goals.

\section*{Acknowledgements and funding}
HH, MG and AB would like to thank Volkswagenstiftung for its support of the "Life?" program (96732).  HH has received funding from the Bundes Ministeriums für Bildung und Forschung under grant agreement No. 031L0237C (MiEDGE project/ ERACOSYSMED). Finally, HH acknowledges the support of the RIG-2023-051 grant from Khalifa University and the UAE-NIH Collaborative Research grant AJF-NIH-25-KU. AB would like to thank Tata Institute of Fundamental Research Hyderabad for the
fellowship.
\section*{Declaration of competing interests}
The authors declare no competing interests.
\section*{Data availability}
This paper does not report either data generation or analysis.

\printbibliography
\newpage
\section*{Supporting material}
\subsubsection*{Polar order}
\begin{figure}[ht!]
    \centering
    \includegraphics[width=0.9\linewidth]{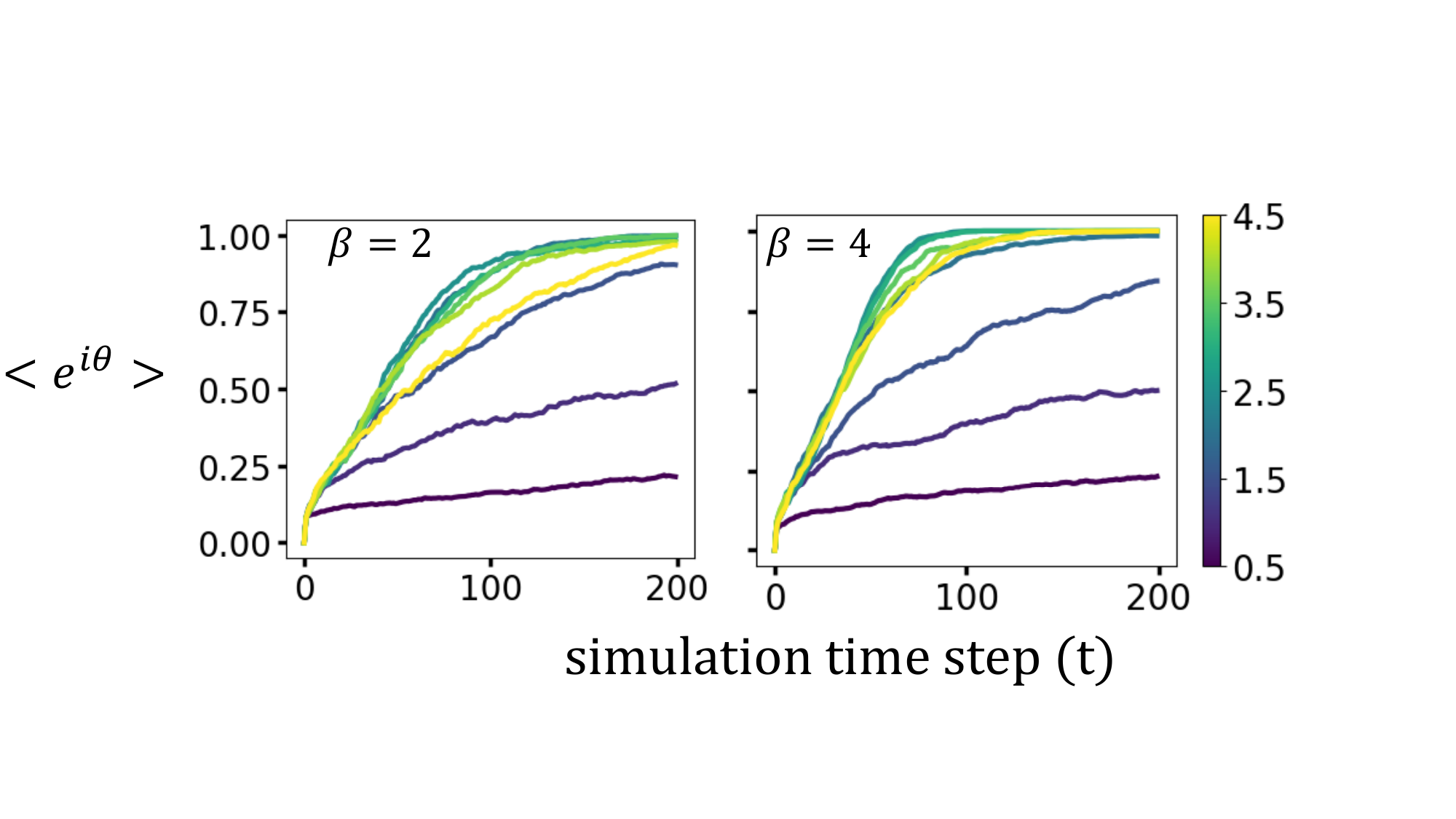}
    \caption{\textbf{ Polar order of the system averaged over 20 simulations under the random choice-making strategy.}}
    \label{polarOrder_random}
\end{figure}

\begin{figure}[ht!]
    \centering
    \includegraphics[width=1\linewidth]{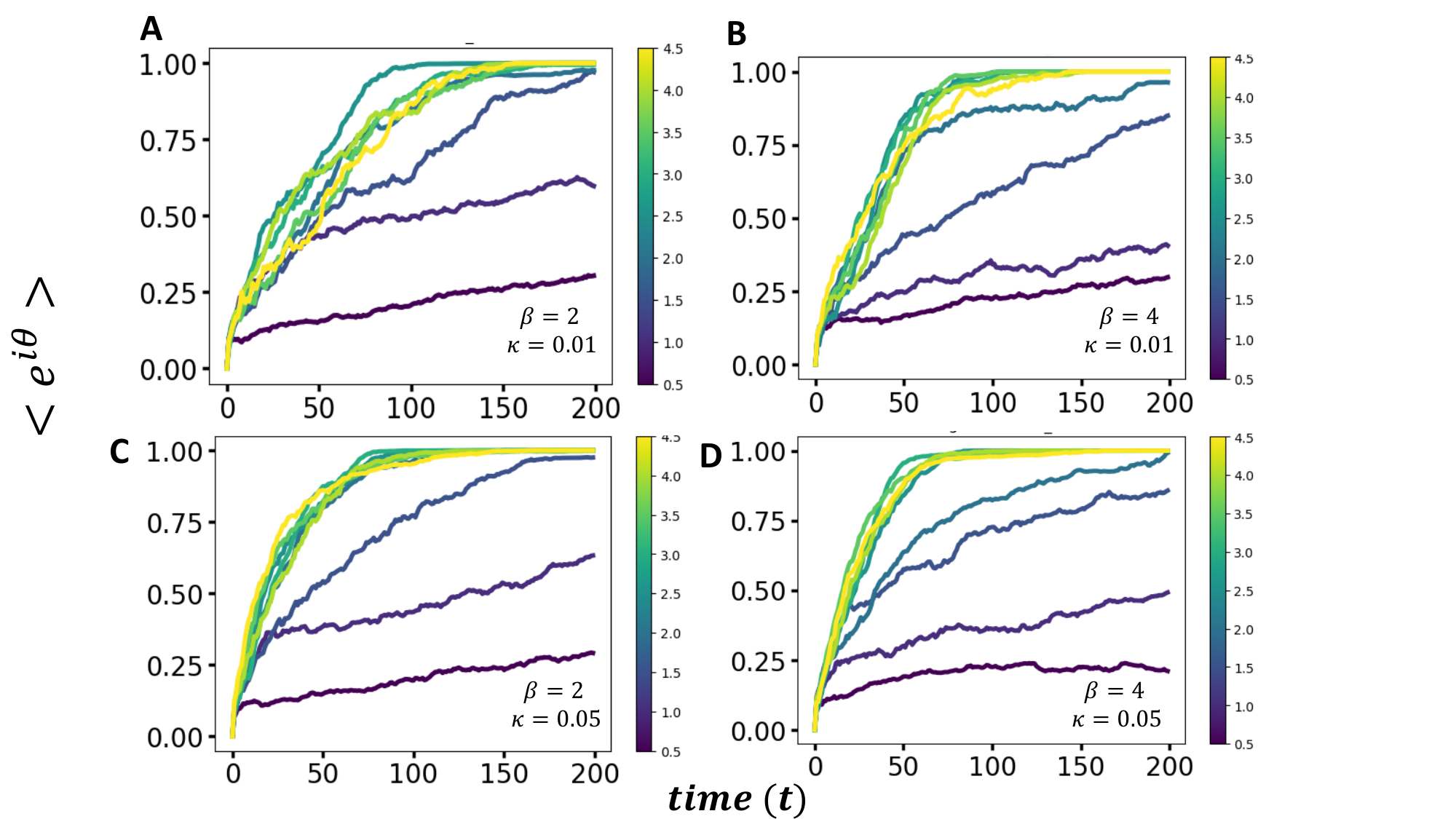}
    \caption{\textbf{Polar order of the system averaged over 20 simulations under the biased strategy.}}
    \label{polarOrder_vonMises}
\end{figure}

\begin{figure}[ht!]
    \centering
    \includegraphics[width=0.5\linewidth]{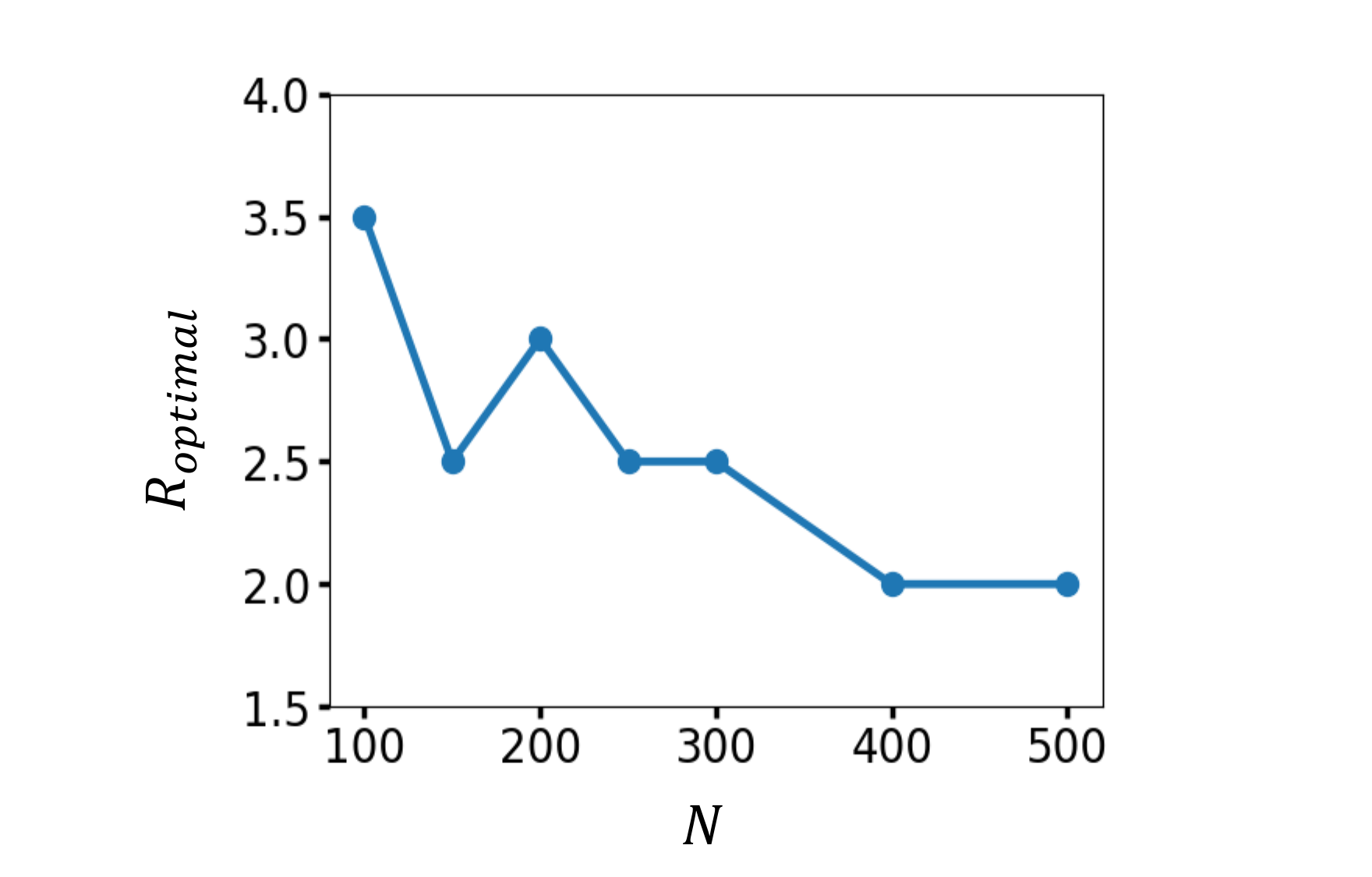}
    \caption{\textbf{Optimal Radius vs Number of agents} As we talked before there is an optimal radius at which synchroniation is fastest. As we increase the radius it increases the number of decision-makers in the neighborhood. Expectedly increasing the density of the agents decreases the optimal radius.}
    \label{optimalRadiusWithN}
\end{figure}
\nopagebreak
\subsubsection*{Analytical proof for the emergence of non-monotonic behavior of the variance of decisions with respect to synchroniation time and radius}

\textbf{1. Derivation of sample variance in terms of population variance and size of the sample} \linebreak
\\
In this section, we shall try to understand why we can observe the paradox of choices in decision-making. Let's consider the microenvironment consists of $N$ decision makers and at the center $n$-th decision maker is collecting all the microenvironmental decision making in a \textit{Gaussian} manner at time $t$ where the mean value of every decision collected from the environment is defined by $\mu_{n}$ and the variance is defined by $\sigma_{n}^{2}$. Here we assumed for simplicity that the angular decisions lies on the interval $(0,2\pi)$. Furthermore, to evaluate average decision-making at the group level we assume a mean-field approach where we use the Central Limit Theorem (CLT) to neglect the correlation among the population. So, the mean value and standard deviation of decision maker $n$ for each local environment can be written (using CLT) as
\begin{equation}
\mu_n =\mu + \frac{\xi}{\sqrt{N}}
\end{equation} 
\begin{equation}
\frac{1}{N}\sum_{i=1}^{N} \left(\mu_n  -\theta_i\right)^2=\frac{1}{N}\sum_{i=1}^{N} \left(\mu -\theta_i +\frac{\xi}{\sqrt{N}}\right)^2
\end{equation} 
\begin{equation}
\sigma^2_n=\frac{1}{N}\sum_{i=1}^{N} \left(\left(\mu -\theta_i\right)^2 + \frac{\xi^2}{N} +\frac{2}{N}(\mu -\sum_{i}\frac{\theta_i}{N})\frac{\xi}{\sqrt{N}}\right)
\end{equation} 
Please note that $\xi$ is the Gaussian noise $\mathcal{N}(0,\Sigma)$ and $\mu$ is the population average of all decision-making in the system. Now, we shall calculate the variance of $n$-th decision maker to do that we shall take the average over the decision-making $\theta$ present in the environment $N$,
%\begin{equation}
%\sigma^2_n=\sum_{i=1}^{N} <((\mu -\theta_i)^2 )>+ (\frac{\Sigma^2\xi^2}{N}) +2(\mu -\sum_{i}\frac{\theta_i}{N})\frac{\Sigma\xi}{\sqrt{N}})^2
%\end{equation} 
%\begin{equation}
%\sigma^2_n=\sum_{i=1}^{N} <((\mu -\theta_i)^2 )>+ \frac{<\xi^2>}{N} +2(\frac{\xi}{\sqrt{N}})\frac{\xi}{\sqrt{N}}
%\end{equation}
%\begin{equation}
%\sigma^2_n=\sum_{i=1}^{N} <((\mu -\theta_i)^2 )>+ \frac{<\xi^2>}{N} +2\frac{<\xi^2>}{N}
%\end{equation}

\begin{equation}
\sigma^2_n=\Sigma + \frac{\Sigma}{N}=\Sigma\left(1+\frac{1}{N}\right)
\label{samplevariance}
\end{equation}
where, $\Sigma$ is the population variance of all decision making in the system. \\

\textbf{2. Derivation of the rate of change in the mean value decisions at the population level}
\\\\
To calculate the evolution of the mean value at the population level we shall use Eq.(\ref{langevin}) and assume the Gaussian form of the microenvironmental entropy. So, the rate of change of decisions can be written as 
\begin{equation}
    \frac{d\theta_n}{dt}=\frac{\beta}{\sigma^{2}_{n}}\left({\frac{\theta_n}{N} -\frac{\sum_{i\neq n}^{N} \theta_i}{N^2}}\right)
\end{equation} 
\begin{equation}
    \frac{d\theta_n}{dt}=\frac{\beta}{\sigma^{2}_n}\left({\frac{\theta_n}{N} -\frac{\sum_{i\neq n}^{N} \theta_i +\theta_n-\theta_n}{N^2}}\right)
\end{equation}   
\begin{equation}
    \frac{d\theta_n}{dt}=\frac{\beta}{\sigma^{2}_{n}}\left({\frac{\theta_n}{N} -\frac{\mu_n}{N}+\frac{\theta_n}{N^2}}\right)
\end{equation}
here, $S_{n}$, and $\mu_n$ are the variance and mean of the sample respectively. Under the central limit theorem, the mean and variance of the sample can be written in terms of the population mean and variance.
\begin{equation}
\mu_n= \mu +\frac{\xi}{\sqrt{N}} \And \sigma^{2}_{n} =\Sigma\left(1+\frac{1}{N}\right)
\end{equation} 
Therefore under the mean-field limit the rate of change of one particle represents a change of the average particle, hence:
\begin{equation}
    \frac{d\theta_n}{dt}=\frac{N}{N+1}\frac{\beta}{\Sigma}\left({\frac{\theta_n}{N} -\frac{\mu+\frac{\xi}{\sqrt{N}}}{N}+\frac{\theta_n}{N^2}}\right)
\end{equation}
\begin{equation}
    \frac{d\langle \theta_n\rangle}{dt}=\frac{N}{N+1}\frac{\beta}{\Sigma}\left({\frac{\langle \theta_n\rangle}{N} -\frac{\mu+\frac{\xi}{\sqrt{N}}}{N}+\frac{\langle \theta_n \rangle}{N^2}}\right)
\end{equation}
\begin{equation}
    \frac{d\mu}{dt}=\frac{N}{N+1}\frac{\beta}{\Sigma}\left({\frac{\mu}{N} -\frac{\mu+\frac{\xi}{\sqrt{N}}}{N}+\frac{\mu}{N^2}}\right)  = \frac{1}{N+1}\frac{\beta}{\Sigma}\left(\frac{\mu}{N}-\frac{\xi}{\sqrt{N}}\right)
\end{equation}
\\
\textbf{3. Derivation of the rate of change of variance of the orientation of the population} \\

Similarly, we shall calculate the evolution of the variance at the population level using the mean-field approach. Using the previous dynamical equation of decisions one can write  
\begin{equation}
\frac{d\Sigma}{dt}=D-\beta\frac{\partial}{\partial \theta_n}\left(\frac{1}{\sigma^{2}_{n}}\left({\frac{\theta_n}{N} -\frac{\sum_{i\neq n}^{N} \theta_i +\theta_n-\theta_n}{N^2}}\right)\right)\Sigma=D+2R\Sigma
\end{equation}
The variance of the random variable has been defined by $\Sigma$ the variance of the Gaussian noise is defined by $D$ and the $R$ term can be seen as
\begin{equation}
    \frac{df}{d\theta_n}=\frac{1}{\sigma^{2}_{n} N}-\frac{\theta_n}{N (\sigma_{n}^2)^2}\frac{d\sigma^{2}_{n}}{d\theta_n}+\frac{\mu_n}{N (\sigma_{n}^2)^2}\frac{d\sigma^{2}_{n}}{d\theta_n}-\frac{\theta_n}{(\sigma_{n}^2)^2N^2}\frac{d\sigma^{2}_{n}}{d\theta_n}
\end{equation}
If we substitute the derivative of variance as $\frac{d\sigma^{2}_{n}}{d\theta_n}$ with $\frac{2(\theta_n-\mu_n)}{N}$
it looks as 
\begin{equation}
    \frac{df}{d\theta_n}=\frac{1}{\sigma^{2}_{n} N}-\frac{2\theta_n(\theta_n-\mu_n)}{N^2 (\sigma_{n}^2)^2}+\frac{2\mu_n (\theta_n-\mu_n)}{N^2 (\sigma_{n}^2)^2}-\frac{2\theta_n(\theta_n-\mu_n)}{(\sigma_{n}^2)^2N^3}
\end{equation}

\begin{equation}
    \frac{df}{d\theta_n}=\frac{1}{\sigma^{2}_{n} N}-\frac{2(\theta_n^2-\theta_n\mu_n)}{N^2(\sigma_{n}^2)^2}+\frac{2 (\theta_n\mu_n-\mu_n^2)}{N^2(\sigma_{n}^2)^2}-\frac{2(\theta_n^2-\mu_n \theta_n)}{(\sigma_{n}^2)^2N^3}
\end{equation}
Furthermore, we replace the term $\mu_n$ with $\mu+\frac{\Sigma}{\sqrt{N}}$ (using CLT) which further simplifies the expression of $\frac{df}{d\theta_n}$ as

\begin{equation}
    \frac{df}{d\theta_n}=\frac{1}{\sigma^{2}_{n} N}-\frac{2(\theta_n^2-\mu \theta_n + \theta_n\frac{\xi}{\sqrt{N}})}{N^2 (\sigma_{n}^2)^2}+\frac{2 (\theta_n\mu + \frac{\theta_n\xi}{\sqrt{N}}-(\mu + \frac{\xi}{\sqrt{N}})^2)}{N^2 (\sigma_{n}^2)^2}
    -\frac{2(\theta_n^2-(\mu \theta_n + \frac{\theta_n\xi}{\sqrt{N}}))}{(\sigma_{n}^2)^2N^3}
\end{equation}
taking an average over the population,
\begin{equation}
\frac{df}{d\theta_{n}}=\frac{1}{\sigma^{2}_{n}N}-\frac{2(\langle \theta_{n}^{2} \rangle-\mu^{2} )}{N^{2} (\sigma_{n}^2)^{2}}+\frac{2 (\mu^2 -\mu^{2} - \frac{\langle\xi^{2}\rangle}{N})}{N^2 (\sigma_{n}^{2})^{2}}-\frac{2(\langle\theta_n^2\rangle-\mu^2)}{(\sigma_{n}^2)^{2}N^{3}}
\end{equation}
in the final step we do a population average as $\langle\theta_n^2\rangle-\mu^2 = \Sigma$ and $\langle\xi^2\rangle$= $\Sigma$

\begin{equation}
    \frac{df}{d\theta_n}=\frac{1}{\sigma^{2}_{n} N}-\frac{2\Sigma}{N^2(\sigma_{n}^2)^2}-
    \frac{2 (\frac{\Sigma}{N})}{N^2 (\sigma_{n}^2)^2}
    -\frac{2\Sigma}{(\sigma_{n}^2)^2N^3}
\end{equation}
now, from Eq.(\ref{samplevariance}), finally replace the $\sigma^{2}_{n}$ as $\Sigma(1+\frac{1}{N})$ we get
\begin{equation}
    \frac{df}{d\theta_n}=\frac{1}{\Sigma(1+ N)}-\frac{2\Sigma}{\Sigma^2(1+N)^2}-
    \frac{2 (\frac{\Sigma}{N})}{\Sigma^2 (1+N)^2}
    -\frac{2\Sigma}{\Sigma^2N(1+N)^2}
\end{equation}
and as we know that from the mean-field equation \cite{Oku2018} one can compare the above expression and can write the variance evolution in the equation zero noise case using eq. $\frac{d\Sigma}{dt}=-2\beta R(\Sigma,\mu,N) \Sigma $   as
\begin{equation}
\frac{d\Sigma}{dt}=-\beta\left(\frac{2}{(1+ N)}-\frac{4}{(1+N)^2}-
    \frac{8}{N(1+N)^2}\right)=-\frac{2\beta\left(N^2-N-4\right)}{N(1+N)^2}
\end{equation}

\end{document}